\documentclass[pra,twocolumn,groupedaddress,showpacs]{revtex4-1}
\usepackage{graphics}

\usepackage{amsmath}
\begin{document}
\newcommand{\op}{\boldsymbol}
\bibliographystyle{apsrev}

\title{Three-Slit Ghost Interference and Nonlocal Duality}

\author{Mohd Asad Siddiqui}
\email{asad@ctp-jamia.res.in}
\affiliation{Centre for Theoretical Physics, Jamia Millia Islamia, New Delhi, India.}


\begin{abstract}
A three-slit ghost interference experiment with entangled photons  is theoretically analyzed using wave-packet dynamics. A {\em nonlocal duality relation} is derived which connects the path distinguishability of one photon to the interference visibility
of the other.

\end{abstract}
\keywords{Entanglement; Nonlocality; Ghost interference.}
\maketitle

\section{Introduction}

Quantum entanglement and nonlocality are two aspects of correlations which are intimately related to each other\cite{nielson}. Such fundamental aspects of quantum theory are extensively studied\cite{horodecki} and today also its an emerging field of research.
The correlated properties of entangled two-photon states have attracted attentions, due to their extensive applications in quantum optics and quantum information\cite{streak,arthur}. As a result, Strekalov et.al  demonstrated the ghost interference experiment\cite{ghostexp}, which show a nonlocal behaviour  with spontaneous parametric down-conversion(SPDC)\cite{kly} source S, a common method of producing entangled photons, conventionally called signal and idler beam\cite{kwait,imaging,walborn}, are then split by a polarized beam splitter into two beams, detected in coincidence by two distant pointlike photon detectors $D_1$ and $D_2$.
 
\begin{figure}[pb]
\centerline{\resizebox{8.4cm}{!}{\includegraphics{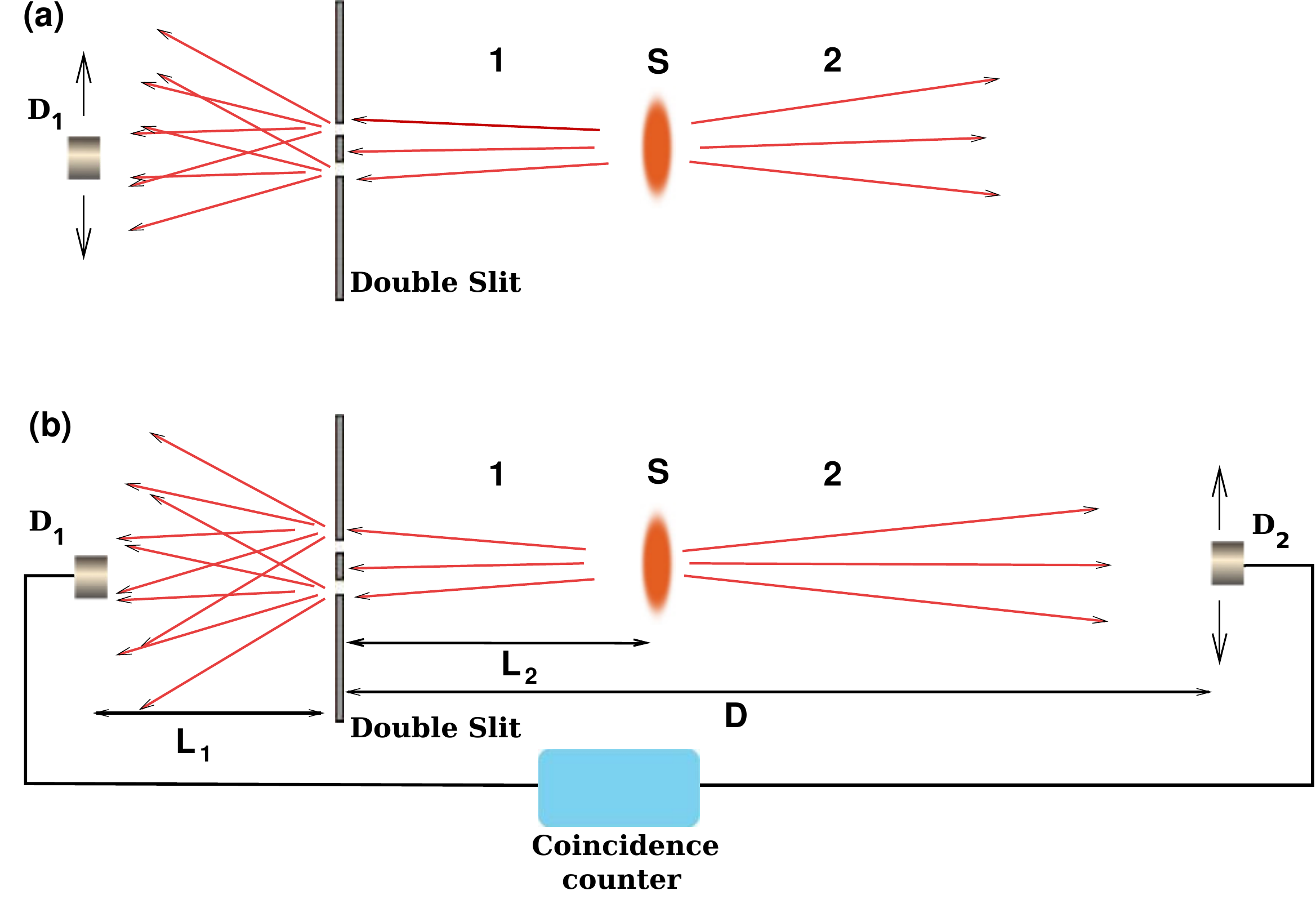}} }
\vspace*{8pt}
\caption{Schematic diagram of the two-slit ghost interference experiment.}
\label{ghostexp}
\end{figure}
A double-slit is in the path of photon 1, and the detector $D_1$ is kept behind (see FIG. \ref{ghostexp}(a)), no  interference pattern is observed for photon 1, surprisingly, as one would normally  expect Young's double-slit interference. Also when the photon 2 is detected by $D_2$,
{\em in coincidence} with a {\em fixed} detector $D_1$,
 the double-slit interference pattern is observed (see FIG. \ref{ghostexp}(b)), even though there is no double-slit in
the path of photon 2. Many interesting outcomes are due to the spatial correlations which is with twin photons, produced in parametric down-conversion\cite{wal}.
 
 The two slit experiment has also been studied extensively in context of wave-particle duality and Bohr's principle of complementarity .
 The fact that the wave and particle nature cannot be observed at the same time, is so fundamental that Bohr gave the
principle, known as, the principle of complementarity\cite{bohr}. Bohr stressed that the wave nature of particle, characterized by interference, and the
particle nature, characterized by which way (i.e., which path)
 information, are mutually exclusive. A further question investigation was if the two natures could be observed simultaneously, and to what level of accuracy. A bound on simultaneous path distinguishability and fringe visibility is described by the so-called Englert-Greenberger-Yasin (EGY) relation\cite{green,englert}.
The EGY duality relation is {\em local}, in the sense that when we talk of which-path distinguishability, we talk of the which-path 
knowledge of the same particle giving interference pattern. A {\em nonlocal} duality relation was derived for two slit experiment\cite{tabish}, which relates the which-path information of one particle to the fringe visibility of the other.

 At present the search for an analogous form of duality relation for  multi slit experiments has generated quite a lot of research activity. Several attempts have been made to explore it quantatively\cite{durr,luis,bim,jak,eng,zaw}. The simplest multi slit case is the three slit case, recently, a duality relation for three slit interference  has been formulated \cite{asad}, a step towards the search for an analogous form of duality relation for more than two slits. The analysis for four or multi slit interference is much more involved than that for two or three slit experiments, there it would be difficult to find  phases for which extreme intensities occur, and thus the visibility.
 
Of late, a focussed interest has been generated towards the three-slit experiments\cite{asad,zel,zaw,urbasi,ste,udu,hess,nies,sawan,gag,guti,gutie}. Three-slits are also used in generating qutrit states, their applications include  implementation  of quantum games\cite{kole}  and in quantum tomography\cite{tagu,pime}.

In this paper, we propose and theoretically analyze a three-slit ghost interference experiment performed with entangled particles. Also a {\em nonlocal} duality relation is derived  which connects the path distinguishability of one particle to the interference visibility of the other. Our analysis is general enough to describe any two entangled particles. It equally well applies to entangled photons, whether generated by {\em SPDC} or any other method like {\em four-wave mixing}.

\section{Three Slit Ghost Interference}

\begin{figure}
\centerline{\resizebox{8.4cm}{!}{\includegraphics{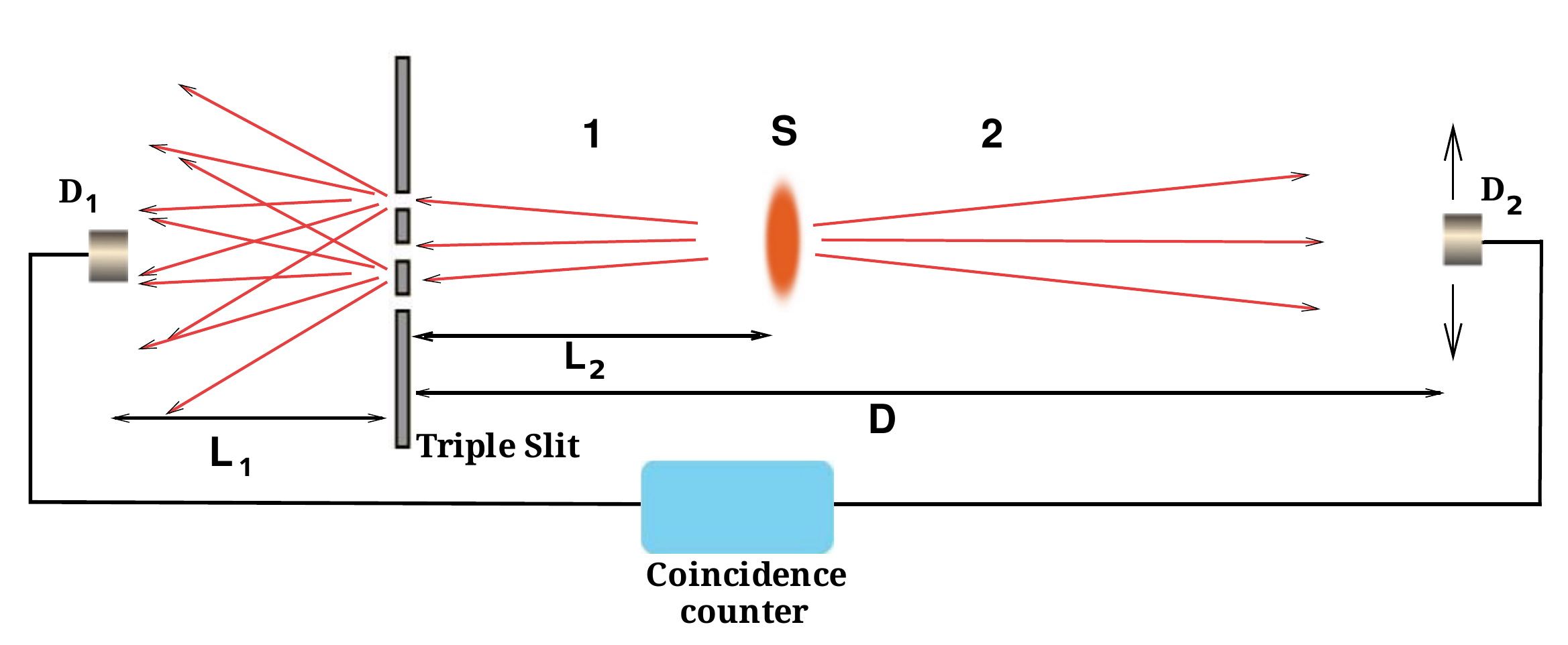}} }
\vspace*{8pt}
\caption{Schematic diagram of the three-slit ghost interference experiment.}
\label{3ghostexpt}
\end{figure}

In our proposed experiment, the {\em two slits} are replaced by {\em three slits}, in the earlier setup  (see FIG. \ref{3ghostexpt}). The entangled photons 1 and 2  from the source S show an interference pattern, similar to the pattern observed from three slits experiment. Even though photon 2 never passes through the region between the source S and three slits, we see an interference pattern for photon 2, as if a beam of photon
2 with a source located at the position of detector $D_1$, get split by three-slits. This behavior can be qualitatively understood with the help of an advanced wave picture introduced by Klyshko\cite{klyshko}. In this picture, the  detector $D_1$ plays the role of a source, which sends photon back towards the crystal. These photons are then reflected by the crystal as by mirror, compels them to follow the path of the signal photons, detected by the detector $D_2$. In this way, the increase in spatial filtering of the detector $D_1$, reduces the size of the source in the advanced wave picture, which increases the spatial coherence. In the following we do a more quantitative analysis.

\section{Wave-Packet Analysis}
Our view is that the ghost interference\cite{tqpcs} is a result of position and momentum entanglement in photon pairs. Same phenomenon should be observed for any two entangled particles. SPDC is just one  method of producing entangled particles. We will base our analysis on entangled pairs of particles.
In order to theoretically analyze the entangled photons, a generalized EPR state\cite{tqajp} is used, which unlike the EPR state\cite{epr}, is well behaved and fully
normalized.
\begin{equation}
\Psi(z_1,z_2) = C_1\!\int_{-\infty}^\infty dp~
e^{-p^2/4\hbar^2\sigma^2}~e^{-ipz_2/\hbar}~ e^{i pz_1/\hbar}~
e^{-{(z_1+z_2)^2\over 4\Omega^2}}, \label{state}
\end{equation}
where $C_1$ is a normalization constant, and $\sigma,\Omega$ are certain
parameters whose physical significance will become clear in the following. In the limit $\sigma,\Omega\to \infty$ the state (\ref{state})
reduces to the EPR state.

The  pair of photons  are assumed to  travel in opposite directions along the x-axis, and the entanglement is in the z-direction. We will ignore the dynamics along the x-axis as it does not affect the entanglement.
We assume that during the evolution for time $t$, the photon travels a
distance equal to $ct$. Integration performed over $p$  in Eq.(\ref{state}) gives:
\begin{equation}
\Psi(z_1,z_2) = \sqrt{ {\sigma\over \pi\Omega}}~
 e^{-(z_1-z_2)^2\sigma^2} e^{-(z_1+z_2)^2/4\Omega^2} .
\label{psi0}
\end{equation}
The uncertainty in positions and the wave-vector of two photons,
along the z-axis, is given by
\begin{eqnarray}
\Delta z_1 &=& \Delta z_2 = \sqrt{\Omega^2+1/4\sigma^2},
\nonumber \\ 
\Delta k_{1z} &=& \Delta k_{2z} = 
{1\over 2}\sqrt{\sigma^2 + {1\over 4\Omega^2}}. \label{unc}
\end{eqnarray}
The above equation gives the position
and momentum spread of the photons in the z-direction.
The time evolution of wave-function is essentially dictated by time evolution of wave-packet.

 If the wave-function of a single photon at time $t=0$ is $\psi(z,0)$, then the wave-function of photon, after time t will evolve as
 \begin{equation}
\psi(z,t) = {1\over 2\pi}\int_{-\infty}^{\infty}
\exp(ik_zz - i\omega(k_z)t)~ \tilde{\psi}(k_z,0)~ dk_z ,
\label{psitl}
\end{equation}
where $\tilde{\psi}(k_z,0)$ is the Fourier transform of $\psi(z,0)$ with
respect to $z$. 

In the above equation, if $k_z=k_0$, then it would be monochromatic approximation. But we have applied an alternative approach.
The photon approximately travels in the x-direction, but can slightly deviate in the z-direction, so it can pass through slits which are located at different z-positions, and therefore its true wave-vector will be given by,
\begin{equation}
\omega(k_z) = c\sqrt{k_x^2 + k_z^2}
\end{equation}
Since the photon travel along x-axis, hence for $k_x \gg k_z$ , one can write
$k_x \approx k_0$, where $k_0$ is the wave-number of the photon associated
with its wavelength, $k_0 = 2\pi/\lambda$. The dispersion along z-axis can be approximated by
\begin{equation}\label{ps}
\omega(k_z) \approx ck_0 + ck_z^2/2k_0.
\end{equation}
The above relation can also be obtained using paraxial approximation, for small angle $\theta$, with $k_x = k_0 \cos\theta$, and $k_z = k_0 \sin\theta$.

Using (\ref{ps}), the Eq.(\ref{psitl}) becomes
\begin{equation}
\psi(z,t) = {e^{-ick_0t}\over 2\pi}\int_{-\infty}^{\infty}
\exp(ik_zz - ictk_z^2/2k_0)~ \tilde{\psi}(k_z,0) ~dk_z
\label{psitk}
\end{equation}

In case of entangled photons, after time $t_0$, photon 1 reaches the triple slit ($ct_0 = L_2$), and photon 2 travels a distance $L_2$ towards detector $D_2$.
Therefore, the wave-function of the entangled photons after  time $t_0$ is given by:
\begin{eqnarray}
\Psi(z_1,z_2,t_0) &=& {e^{-2ick_0t_0}\over 4\pi^2}\int_{-\infty}^{\infty} dk_1
\exp(ik_1z_1 - ict_0k_1^2/2k_0)\nonumber\\
&\times &\int_{-\infty}^{\infty} dk_2
\exp(ik_2z_2 - ict_0k_2^2/2k_0)~ \tilde{\Psi}(k_1,k_2,0),
\nonumber\\ 
\label{psit0}
\end{eqnarray}
where $\tilde{\Psi}(k_1,k_2,0)$ is the Fourier transform of (\ref{psi0}) with
respect to $z_1,z_2$.

To investigate the effect on the entangled state, one can use two different approaches. The first, most obvious is to model a potential for three-slits, and calculate its evolution in that potential. We will follow the second, a comparatively easier approach, here we capture the essence of the effect of triple slit on the wave-function, without going into tedious calculations. When the state interacts with a single-slit, we assume, that a Gaussian wave-packet emerges from that slit, centered at its location, whose width is related to the width of the slit.

Consider the state of particle 1 passing through the slits  A, B and C be $|\phi_A\rangle$, $|\phi_B\rangle$ and $|\phi_C\rangle$, respectively. Some part of the state of particle 1 will be blocked, represented by $|\chi\rangle$. All these states are orthogonal, and the actual state of particle 1 can be expanded in this basis. 
\begin{eqnarray}
|\Psi(t_0)\rangle &=& |\phi_A\rangle\langle\phi_A|\Psi\rangle
+ |\phi_B\rangle\langle\phi_B|\Psi\rangle \nonumber \\ &+&   |\phi_C\rangle\langle\phi_C|\Psi\rangle +
|\chi\rangle\langle\chi|\Psi\rangle . \label{psit1}
\end{eqnarray}
The terms $\langle\phi_A|\Psi\rangle,~ \langle\phi_B|\Psi\rangle,~  \langle\phi_C|\Psi\rangle,~  \langle\chi|\Psi\rangle$ are states of particle 2 and can be written explicitly as follows.
\begin{eqnarray}
|\psi_A \rangle &=& \langle\phi_A |\Psi(t_0)\rangle, \nonumber\\
|\psi_B \rangle &=& \langle\phi_B |\Psi(t_0)\rangle, \nonumber\\
|\psi_C \rangle &=& \langle\phi_C |\Psi(t_0)\rangle, \nonumber\\
|\psi_\chi \rangle &=& \langle\chi |\Psi(t_0)\rangle, \label{psit2}
\end{eqnarray}
The entangled state after particle 1 passes through triple-slit will be given by:
\begin{equation}
|\Psi\rangle = |\phi_A\rangle|\psi_A\rangle
+ |\phi_B\rangle|\psi_B\rangle + |\phi_C\rangle|\psi_C\rangle +
|\chi\rangle|\psi_\chi\rangle ,
\end{equation}
where $|\phi_A\rangle$ , $|\phi_B\rangle$ and $|\phi_C\rangle$ are states of particle 1,
and $|\psi_A\rangle$ , $|\psi_B\rangle$ and $|\psi_C\rangle$ are states of particle 2.
In general, even if $|\phi_A\rangle$, $|\phi_B\rangle$ and $|\phi_C\rangle$ are orthogonal, $|\psi_A\rangle$, $|\psi_B\rangle$ and $|\psi_C\rangle$ may or may not be orthogonal depending on the values of $\Omega$ and $\sigma$, which dictate the degree of correlation between two particles. Perfect correlation will happen only when $\sigma,\Omega\to \infty$ , in that case Eq.(\ref{state})
becomes the idealized EPR state.

The first three term represents the amplitude of particle 1 passing through these slits, and the last term represents the amplitude of  being blocked or reflected. The linearity of the Schr\"odinger equation assures that
the first three terms and the last term evolve independently. Since the experiment consider only those photon 1 which passes through the triple slit, we can throw away the last term. This will not change anything except the renormalization of the state.

For simplicity, we assume that  $\langle z_1|\phi_A \rangle$, $\langle z_1|\phi_B \rangle$, and $\langle z_1|\phi_C \rangle$  are
Gaussian wave-packets:
\begin{eqnarray}
\langle z_1|\phi_A \rangle = \phi_A(z_1) &=& {1\over(\pi/2)^{1/4}\sqrt{\epsilon}} e^{-(z_1-z_0)^2/\epsilon^2}
,\nonumber\\
\langle z_1|\phi_B \rangle = \phi_B(z_1) &=& {1\over(\pi/2)^{1/4}\sqrt{\epsilon}} e^{-z_1^2/\epsilon^2}
,\nonumber\\
\langle z_1|\phi_C \rangle = \phi_C(z_1) &=& {1\over(\pi/2)^{1/4}\sqrt{\epsilon}} e^{-(z_1+z_0)^2/\epsilon^2},\label{psit}
\end{eqnarray}
where $+z_0, 0 ,-z_0 \-$ are z-position's of slit A, B and C, respectively, and $\epsilon$ be
their widths. 

Using (\ref{psit0}), (\ref{psit1}), (\ref{psit2}) and (\ref{psit}), wave-functions  $\langle z_2|\phi_A \rangle$, $\langle z_2|\phi_B \rangle$, and $\langle z_2|\phi_C \rangle$  can be calculated, which, after normalization, has the following form
\begin{eqnarray}
\langle z_2|\phi_A \rangle = \psi_A(z_2) &=& C_2 ~e^{-{(z_2 - z_0')^2 \over \Gamma}},\nonumber\\
\langle z_2|\phi_B \rangle = \psi_B(z_2) &=& C_2 ~e^{-{z_2^2 \over \Gamma}} ,\nonumber\\
\langle z_2|\phi_C \rangle = \psi_C(z_2) &=& C_2 ~e^{-{(z_2 + z_0')^2 \over \Gamma}} ,
\end{eqnarray}
where 
\begin{equation}
\Gamma = \frac{\epsilon^2+ {1\over\sigma^2}+{\epsilon^2\over 4\Omega^2\sigma^2}
 + {2i\hbar t_0\over m} 
}{1 + {\epsilon^2\over\Omega^2}+{i2\hbar t_0\over 4\Omega^2m} +
{1\over 4\Omega^2\sigma^2}} + {2i\hbar t_0\over m}, \nonumber
\end{equation}
\begin{equation}
z_0' = {z_0 \over {4\Omega^2\sigma^2+1\over
4\Omega^2\sigma^2-1} + {4\epsilon^2 \over 4\Omega^2-1/\sigma^2}}, \nonumber
\end{equation}
and
\begin{center}
$C_2 = (2/\pi)^{1/4}(\sqrt{\Gamma_r} + {i\Gamma_i\over\sqrt{\Gamma_r}})^{-1/2}$.
\end{center} 
Here $\Gamma_r, ~\Gamma_i$ are the real and imaginary parts of $\Gamma$,
respectively.

Thus, the wave-function which emerges from the triple slit, has the following form
\begin{eqnarray}
\Psi(z_1,z_2) &=& C~\Big(e^{{-(z_1-z_0)^2\over\epsilon^2}}
e^{{-(z_2 - z_0')^2 \over \Gamma}} 
+ e^{{-z_1^2\over\epsilon^2}}
e^{{-z_2^2 \over \Gamma}} \nonumber \\ &+& 
 e^{{-(z_1+z_0)^2\over\epsilon^2}}
e^{{-(z_2 + z_0')^2 \over \Gamma}}\Big) \label{virtual},
\end{eqnarray}

where $C = (\sqrt{2/3\pi\epsilon})(\sqrt{\Gamma_r} +
{i\Gamma_i\over\sqrt{\Gamma_r}})^{-1/2}$.

 The above expression is obtained by dropping the phase factor of Eq.(\ref{psit0}), as it is not important for
our final analysis.
Eq.(\ref{virtual}) represents three wave-packets of photon 1,
of width $\epsilon$, and localized at $-z_0$, $0$ and $+z_0$, entangled with three
wave-packets of photon 2, of width
${\sqrt{2}|\Gamma|\over\sqrt{\Gamma+\Gamma^*}}$, localized at $-z_0$, $0$ and $+z_0$.

At this  stage one can notice the amplitude of photon 1 through slits A, B and C, which are correlated to spatially separated
wave-packets of 
photon 2. Thus, in principle  one can detect the photon 2, and therefore which
slit, A, B or C, the photon 1 passed through. By Bohr's principle of complementarity,
if one knows which slit  the photon 1 passed through, no
interference pattern will be seen. This is the fundamental reason
for non-observance of interference pattern by photon 1 in the ghost
interference experiment.

Before reaching detector $D_2$, the particle 2 further evolves for  time $t$, thus transforms the wave-function (\ref{virtual}) to
\begin{eqnarray}
\Psi(z_1,z_2,t)&=& C_t~\Bigg(\exp\left[{{-(z_1-z_0)^2\over\epsilon^2 +{iL_1\lambda\over\pi}}}\right]
 \exp\left[{{-(z_2 - z_0')^2 \over \Gamma +{iL_1\lambda\over\pi}}}\right]\nonumber \\ &+& 
 \exp\left[{{-z_1^2\over\epsilon^2 +{iL_1\lambda\over\pi}}}\right]
 \exp\left[{{-z_2^2 \over \Gamma +{iL_1\lambda\over\pi}}}\right]
\nonumber\\ 
&+&  \exp\left[{{-(z_1+z_0)^2\over\epsilon^2+{iL_1\lambda\over\pi}}}\right]
 \exp\left[{{-(z_2 + z_0')^2 \over \Gamma+{iL_1\lambda\over\pi}}}\right]\Bigg),\nonumber\\
\label{psifinal}
\end{eqnarray}
where 
\begin{equation}
C_t = {\sqrt2\over \sqrt{3\pi}\sqrt{\epsilon + iL_1\lambda/\epsilon\pi}
\sqrt{\sqrt{\Gamma_r}+(\Gamma_i +iL_1\lambda/\pi)/\sqrt{\Gamma_r}}}.
\nonumber
\end{equation}

When the correlation because of entanglement between the photons are good, one can make further approximations:
$\Omega \gg \epsilon$, $\Omega \gg 1/\sigma$ and $\Omega \gg 1$. In this
limit,
\begin{equation}
\Gamma \approx \gamma^2 + 2i\hbar t_0/\mu,~~~ z_0' \approx z_0,
\end{equation}
where $\gamma^2 = \epsilon^2 + 1/\sigma^2$.

The wave-function (\ref{psifinal}) represents the combined state of two photons when they reach the detector $D_1$ and $D_2$.
Now if $D_1$ and $D_2$ are located at $z_1$ and $z_2$ respectively, the probability density of their coincident count is given by
\begin{eqnarray}
P(z_1,z_2) &=& |\Psi(z_1,z_2,t)|^2 \nonumber\\
&=& |C_t|^2 \Bigg(\exp\left[-{2(z_1 -z_0)^2\over\epsilon^2+({\lambda L_1\over\pi\epsilon})^2}
-{2(z_2 - z_0)^2 \over \gamma^2+({\lambda D\over\pi\gamma})^2}\right]
\nonumber \\ &+&  \exp\left[-{2 z_1^2\over\epsilon^2+({\lambda L_1\over\pi\epsilon})^2}
-{2 z_2^2 \over \gamma^2+({\lambda D\over\pi\gamma})^2}\right]
\nonumber\\
&+& \exp\left[-{2(z_1 +z_0)^2\over\epsilon^2+({\lambda L_1\over\pi\epsilon})^2}
-{2(z_2 + z_0)^2 \over \gamma^2+({\lambda D\over\pi\gamma})^2}\right]
\nonumber \\ &+&  \exp\left[-{2 z_1^2+z_0^2-2 z_1 z_0\over\epsilon^2+({\lambda L_1\over\pi\epsilon})^2}
-{2 z_2^2 + z_0^2 - 2 z_2 z_0 \over \gamma^2+({\lambda D\over\pi\gamma})^2}\right]\nonumber\\
&\times & 2\cos\left[(z_0^2-2z_1 z_0 )\xi_1 + (z_0^2-2z_2 z_0)\xi_2 \right]
\nonumber \\ &+&  \exp\left[-{2(z_1^2+z_0^2)\over\epsilon^2+({\lambda L_1\over\pi\epsilon})^2}
-{2(z_2^2 + z_0^2) \over \gamma^2+({\lambda D\over\pi\gamma})^2}\right]\nonumber\\
&\times & 2\cos\left[4 z_0 z_1 \xi_1 + 4 z_0 z_2 \xi_2 \right]
\nonumber \\ &+&  \exp\left[-{2 z_1^2+z_0^2+2 z_1 z_0\over\epsilon^2+({\lambda L_1\over\pi\epsilon})^2}
-{2 z_2^2 + z_0^2 + 2 z_2 z_0 \over \gamma^2+({\lambda D\over\pi\gamma})^2}\right]
\nonumber\\
&\times &  2\cos\left[(z_0^2+2z_1 z_0 )\xi_1 + (z_0^2+2z_2 z_0)\xi_2 \right]\Bigg),
\label{pattern}
\end{eqnarray}
where 
\begin{equation}
\xi_1 = {\lambda L_1/\pi\over \epsilon^4 + (\lambda L_1/\pi)^2},~~~
\xi_2 = {\lambda D /\pi\over \gamma^4 + (\lambda D/\pi)^2}, \nonumber
\end{equation}
\begin{center}
$D = L_1 + 2 L_2$,~~~ and
\end{center}
\begin{equation}
C_t = {\sqrt2\over \sqrt{3\pi}\sqrt{\epsilon+{i\lambda L_1\over\pi\epsilon}}
\sqrt{\gamma+{i\lambda D\over\pi\gamma}}}. 
\nonumber
\end{equation}

\section{Results}

\subsection{Ghost interference}

We analyze three slit ghost interference experiment, the
entangled photons with  wave-length $\lambda$, and the detector $D_1$ is fixed at $z_1=0$. In that case, (\ref{pattern}) reduces to
\begin{eqnarray}
|\Psi(0,z_2,t)|^2 &=& |C_t|^2 \Bigg[\exp\left(\frac{-2z_0^2}{\epsilon^2 +(\frac{\lambda L_{1}}{\pi\epsilon})^2}\right)
\nonumber 
\exp\left(\frac{-2(z_2^2+z_0^2)}{\gamma_D^2}\right)
\nonumber \\ &\times & 
 2\cosh\left[\frac{4z_2 z_0}{\gamma_D^2}\right]
\left(1+\frac{\cos\left[\frac{4 z_2 z_0 \pi}{\lambda D} \right]}{\cosh\left[\frac{4z_2 z_0}{\gamma_D^2}\right]}\right)
\nonumber
\\ &+&
\exp\left(\frac{-2z_2^2}{\gamma_D^2}\right) + 2\exp\left(\frac{-z_0^2}{\epsilon^2 +(\frac{\lambda L_{1}}{\pi\epsilon})^2}-\frac{2(z_2^2+z_0^2)}{\gamma_D^2}\right) \nonumber \\ &\times & 
\Bigg(\exp \left(\frac{2z_2 z_0}{\gamma_D^2}\right) \cos\left[\frac{2z_2 z_0 \pi}{\lambda D}-\beta \right]
\nonumber
\\ &+&
\exp \left(\frac{-2z_2 z_0}{\gamma_D^2}\right) \cos\left[\frac{2z_2 z_0 \pi}{\lambda D}+\beta \right]\Bigg)
\Bigg],
\end{eqnarray}

where
\begin{center}
 ~$\gamma_D^2 = \gamma^2 + (\lambda D/\pi\gamma)^2$,~~~~and
\end{center}
\begin{equation}
\beta = {z_0^2 ~\pi \over \lambda}\left(\frac{1}{L_1}+\frac{1}{D}\right). \nonumber
\end{equation}
Neglecting $\beta$ , we get
\begin{eqnarray}
|\Psi(0,z_2,t)|^2 &=& |C_t|^2 \Bigg(\exp\left[\frac{-2z_0^2}{\epsilon^2 +(\frac{\lambda L_{1}}{\pi\epsilon})^2}\right]
\exp\left[\frac{-2(z_2^2+z_0^2)}{\gamma_D^2}\right]
\nonumber \\ &\times &  
 2\cosh\left[\frac{4z_2 z_0}{\gamma_D^2}\right]
\left[1+\frac{\cos\left[\kappa_1 z_2 \right]}{\cosh\left[\frac{4z_2 z_0}{\gamma_D^2}\right]}\right]
\nonumber
\\ &+&
\exp\left[\frac{-2z_2^2}{\gamma_D^2}\right] + 2\exp\left[\frac{-z_0^2}{\epsilon^2 +(\frac{\lambda L_{1}}{\pi\epsilon})^2}-\frac{2(z_2^2+z_0^2)}{\gamma_D^2}\right]
\nonumber
\nonumber \\ &\times & 
2 \cosh \left[\frac{2z_2 z_0}{\gamma_D^2}\right] 
\cos\left[\kappa_2 z_2  \right] \Bigg), 
\label{patteri}
\end{eqnarray}

where 
\begin{equation}
\kappa_1 = {4 \pi z_0 \over \lambda D},~~~
\kappa_2 = {2 \pi z_0 \over \lambda D}, \nonumber
\end{equation}

For $\gamma^2\ll \lambda D/\pi$, 
(\ref{patteri}) represents an interference pattern for photon 2
with fringe widths \Big( $( w_2=\frac{2\pi}{\kappa_i})$, where $ i\in 1,2 $ \Big), due to slit A and C, A and B , and, B and C, are respectively given by, 
\begin{eqnarray}
 (w_2)_{AC} &\approx & {\lambda D\over 2 z_0},\nonumber \\
 (w_2)_{AB} &=& (w_2)_{BC} \approx {\lambda D\over z_0}.
\label{young1}
\end{eqnarray}
This is the ghost interference, the distance $D$ in the formula is the distance from the three-slits, right through the source to the detector $D_2$, (see FIG. \ref{3ghostexpt}).

\subsection{Nonlocal wave-particle duality}

\begin{figure}[pb]
\centerline{\resizebox{8.4cm}{!}{\includegraphics{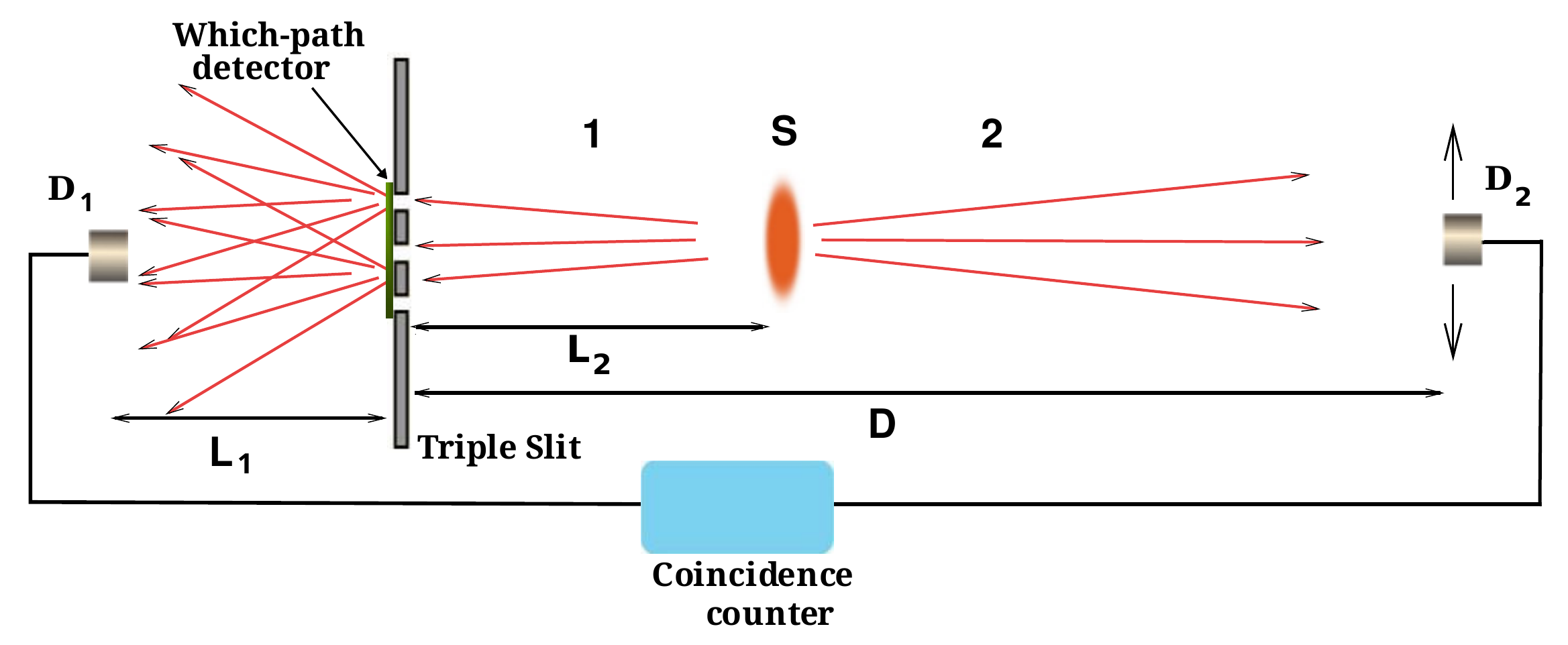}} }
\vspace*{8pt}
\caption{Three-slit ghost interference experiment, when which-way detector is placed behind the slits.}
\label{wghostexpt}
\end{figure}
To find the duality relation, we place the which-way detector behind the three slits, (see FIG. \ref{wghostexpt}), by  which the experimenter gets which-path information. The which-way detector chosen with three states, correlate with the particle, when it passes through each slit.
Let the path-detector states be $|d_1\rangle, |d_2\rangle, |d_3\rangle$, which
correspond to the particle passing through slit 1, 2 and 3, respectively.
Without the loss of generality, we assume that the states 
$|d_1\rangle, |d_2\rangle, |d_3\rangle$ are normalized, but not necessarily mutually orthogonal.

A fundamental property of quantum mechanics is that a state cannot be perfectly distinguished  by any physical device, unless they are orthogonal. However, if a non-zero probability of inconclusive answer is allowed, one can certainly distinguish the given sates. This idea was introduced by Ivanovic\cite{iva}, Dieks\cite{diek} and Peres\cite{pere} and is called unambiguous quantum state discrimination(UQSD). The above strategy can be used to  gain the information about the path taken by the particle in interference experiments.

If path-detector states $|d_1\rangle, |d_2\rangle, |d_3\rangle$ are mutually orthogonal, one can get the information about the path of the particle, without ambiguity. For non-orthogonal states $|d_1\rangle, |d_2\rangle, |d_3\rangle$ we use an UQSD technique to define distinguishability\cite{asad}, for three slit interference.
\begin{equation}
{\mathcal D_Q} \equiv 1 - {1\over 3}(|\langle d_1|d_2\rangle| + 
|\langle d_2|d_3\rangle| +|\langle d_1|d_3\rangle|),
\end{equation}
The which-way {\em distinguishability} for particle 1 is given by
\begin{equation}
{\mathcal D^1_Q} \equiv 1 - {1\over 3}(|\langle d_1|d_2\rangle| + 
|\langle d_2|d_3\rangle| +|\langle d_1|d_3\rangle|),
\label{D}
\end{equation}
the value lies in the range $0\le {\mathcal D^1_Q}\le 1$.

Let us see the effect of  which-path detector on the ghost
interference given by particle 2. We assume that the two particles move in opposite directions along the x-axis, and the entanglement is in the z-direction.

The  particle 1 is then made to interact with which-path detector, which gives rise to an entanglement between the two particles and the  which-path detector.

We get the following states.
\begin{eqnarray}
\Psi(z_1,z_2,t)&=& C_t \Bigg(|d_1\rangle \exp\left[{{-(z_1-z_0)^2\over\epsilon^2 +{iL_1\lambda\over\pi}}}\right]
 \exp\left[{{-(z_2 - z_0')^2 \over \Gamma +{iL_1\lambda\over\pi}}}\right]
\nonumber \\ &+&   |d_2\rangle \exp\left[{{-z_1^2\over\epsilon^2 +{iL_1\lambda\over\pi}}}\right]
 \exp\left[{{-z_2^2 \over \Gamma +{iL_1\lambda\over\pi}}}\right]
\nonumber\\ 
&+& 
|d_3\rangle \exp\left[{{-(z_1+z_0)^2\over\epsilon^2+{iL_1\lambda\over\pi}}}\right]
 \exp\left[{{-(z_2 + z_0')^2 \over \Gamma+{iL_1\lambda\over\pi}}}\right]\Bigg),\nonumber\\
\label{sifinal}
\end{eqnarray}
where 
\begin{equation}
C_t = {\sqrt2\over \sqrt{3\pi}\sqrt{\epsilon + iL_1\lambda/\epsilon\pi}
\sqrt{\sqrt{\Gamma_r}+(\Gamma_i +iL_1\lambda/\pi)/\sqrt{\Gamma_r}}}.\nonumber
\end{equation}

The probability density at $z_1=0$, given by
$|\Psi(0,z_2,t)|^2$, has the following form
\begin{eqnarray}
|\Psi(0,z_2,t)|^2 &=& |C_t|^2 \Bigg[\exp\left[\frac{-2z_2^2}{\gamma_D^2}\right] \Bigg(1 +2\cosh\left[\frac{4z_2 z_0}{\gamma_D^2}\right] \nonumber \\ &\times &  \exp\left[\frac{-2z_0^2}{\epsilon^2 +(\frac{\lambda L_{1}}{\pi\epsilon})^2}-\frac{2 z_0^2}{\gamma_D^2}\right] \Bigg)  + 2~|\langle d_1|d_2\rangle |\nonumber \\ &\times & 
  \exp\left(\frac{-z_0^2}{\epsilon^2 +(\frac{\lambda L_{1}}{\pi\epsilon})^2} -\frac{2 z_2^2+z_0^2-2z_2 z_0 }{\gamma_D^2}\right) \nonumber \\ &\times & 
\cos\left[2z_2 z_0 \xi_2 - z_0^2 (\xi_1 + \xi_2) \right]
\nonumber
\\ &+& 2~|\langle d_1|d_3\rangle | \exp\left(\frac{-2z_0^2}{\epsilon^2 +(\frac{\lambda L_{1}}{\pi\epsilon})^2}-\frac{2(z_2^2+z_0^2)}{\gamma_D^2}\right)\nonumber \\ &\times & 
 \cos\left[4z_2 z_0 \xi_2 \right] \nonumber\\ &+&
2~|\langle d_2|d_3\rangle |
\exp\left(\frac{-z_0^2}{\epsilon^2 +(\frac{\lambda L_{1}}{\pi\epsilon})^2} -\frac{2 z_2^2+z_0^2+2z_2 z_0 }{\gamma_D^2}\right)\nonumber \\ &\times & 
\cos\left[2z_2 z_0  \xi_2 + z_0^2 (\xi_1 + \xi_2)  \right]
\Bigg],\nonumber\\
\label{peter}
\end{eqnarray}
where 
\begin{equation}
\xi_1 = {\lambda L_1/\pi\over \epsilon^4 + (\lambda L_1/\pi)^2},~~~
\xi_2 = {\lambda D /\pi\over \gamma^4 + (\lambda D/\pi)^2}, \nonumber
\end{equation}

Visibility of the interference fringes is conventionally defined as\cite{born}
\begin{equation}
{\mathcal V} = {I_{max} - I_{min} \over I_{max} + I_{min} } ,
\end{equation}
where $I_{max}$ and $I_{min}$ represent the maximum and minimum intensity
in neighbouring fringes, respectively.
Maxima and minima of (\ref{peter}) will occur at points where the 
value of each cosine is 1 and -1/2 , respectively, provided we ignore $z_0^2(\xi_1+\xi_2)$ term. If we look at any fringe, other than the central one, $z_0 \ll z_2$, and hence can be ignored in comparison.

The visibility of particle 2 can then be written down as
\begin{equation}
{\mathcal V}_2 = {3\left( |\langle d_1|d_2\rangle| e^{{2z_2 z_0\over \gamma_D^2}}+ |\langle d_1|d_3\rangle| e^{-z_0^2 \zeta}+ |\langle d_2|d_3\rangle| e^{{-2z_2 z_0\over \gamma_D^2}}\right) \over \alpha +|\langle d_1|d_2\rangle| e^{{2z_2 z_0\over \gamma_D^2}}+ |\langle d_1|d_3\rangle| e^{-z_0^2 \zeta}+ |\langle d_2|d_3\rangle| e^{{-2z_2 z_0\over \gamma_D^2}} }.
\end{equation}

\begin{center}
where, ~~~~$\zeta = \left[\frac{1}{\epsilon^2 +(\frac{\lambda L_{1}}{\pi\epsilon})^2}+\frac{1}{\gamma_D^2}\right],$ ~~~and \\
$\alpha = 2 \left(\exp[\frac{z_0^2 }{\gamma_D^2}] + 2 \exp[-{z_0^2\zeta}] \cosh[\frac{4 z_2 z_0 }{\gamma_D^2} ]\right)$,
\end{center}
The maximum visibility one can theoretically get  when $ z_0 \ll \frac{\lambda L_1}{\pi}$, and $z_0 \ll \frac{\lambda D}{\pi} $. The actual fringe visibility will be
less than or equal to that, and can be written as
\begin{equation}
{\mathcal V_2} \le {3\left( |\langle d_1|d_2\rangle| + |\langle d_1|d_3\rangle| + |\langle d_2|d_3\rangle| \right) \over 2 \left( 1 + 2 \right)+|\langle d_1|d_2\rangle| + |\langle d_1|d_3\rangle| + |\langle d_2|d_3\rangle| }.
\end{equation}
Using (\ref{D}), the above equation gives 
\begin{equation}
{\mathcal V_2} + {2{\mathcal D^1_Q}\over 3- {\mathcal D^1_Q}} \le 1 .
\label{duality}
\end{equation}

The duality relation (\ref{duality}),  is very similar to the duality relation derived earlier for a three-slit interference experiment\cite{asad}. The big difference is that, in three-slit experiment we talk of the  path distinguishability and  the fringe visibilty for the  same particle. In three-slit ghost interference, we show that the relation is between different particles, i.e the path distinguishability  of particle 1 is related with the fringe visibilty of particle 2.

If instead of triple slit, a double slit were kept in the path, the path distinguishability  of particle 1 and the fringe visibilty of particle 2, will follow a different duality relation, given by ${\mathcal V_2} + {\mathcal D^1_Q} \le 1$. This can be inferred by relating $D_Q$ with distinguishability used in Ref. [\cite{tabish}].

\section{Conclusion}
In conclusion, we have analyzed the complementarity  between which-way information and interference fringe visibility for the ghost interference, for  entangled photons passing through three slits. We also derive a three-slit nonlocal duality relation  which connects the path distinguishability of one photon to the interference visibility of the other which means  erasing the which-path information of  photon 1 recovers
the interference pattern of photon 2 and vice-versa.

\section*{Acknowledgments}

M.A. Siddiqui acknowledges financial support from UGC, India and he
thanks Tabish Qureshi for useful discussions.

\end{document}